# Dynamic Kernel Matching for Non-conforming Data:
# A Case Study of T-cell Receptor Datasets

**Jared Ostmeyer**[*]                                    Jared.Ostmeyer@UTSouthwestern.edu
**Scott Christley**
**Lindsay Cowell**
*Department of Population and Data Sciences*
*University of Texas Southwestern Medical Center*
*Dallas, Texas, United States of America*

## Abstract

Most statistical classifiers are designed to find patterns in data where numbers fit into rows and columns, like in a spreadsheet, but many kinds of data do not conform to this structure. To uncover patterns in non-conforming data, we describe an approach for modifying established statistical classifiers to handle non-conforming data, which we call *dynamic kernel matching* (DKM). As examples of non-conforming data, we consider (i) a dataset of T-cell receptor (TCR) sequences labelled by disease antigen and (ii) a dataset of sequenced TCR repertoires labelled by patient cytomegalovirus (CMV) serostatus, anticipating that both datasets contain signatures for diagnosing disease. We successfully fit statistical classifiers augmented with DKM to both datasets and report the performance on holdout data using standard metrics and metrics allowing for indeterminant diagnoses. Finally, we identify the patterns used by our statistical classifiers to generate predictions and show that these patterns agree with observations from experimental studies.

Source code at https://github.com/jostmey/dkm

## Introduction

Statistical classifiers are mathematical models that use example data to find patterns in features that predict a label. The features and corresponding label are referred to as a sample. Standard statistical classifiers implicitly assume (i) each sample has the same number of features and (ii) each feature has a fixed input occupying the same position across all samples (Fig. 1a). For example, the first feature might always represent a patient's age and the second feature might always represent their height. Under these assumptions, each input of the statistical classifier always gets features conveying the same kinds of information. We call features non-conforming when these assumptions do not hold, requiring specialized approaches to handle a lack of correspondence between the non-conforming features and the inputs of the statistical classifier (Fig. 1b).

Sequences are one example of a datatype with non-conforming features. The essential property of a sequence is that both the content and the order of the symbols in the sequence convey information. Sequence data is non-conforming because some sequences are longer than others, resulting in irregular numbers of features. Even when sequences are the same length, a pattern of symbols shared between these sequences can appear at different positions, preventing the same feature from appearing at the same position across all samples. To handle sequences, statistical classifiers need to be able to find patterns in either the content or the order of the symbols that predict the label. Shimodaira and colleagues developed an approach that does both [1]. Their statistical classifier uses weights to both tune how each feature contributes to the prediction and place each feature into context based on its order in the sequence. Their approach, originally developed for support vector machines, has been applied to deep neural networks by Iwana and Uchida [2, 3].

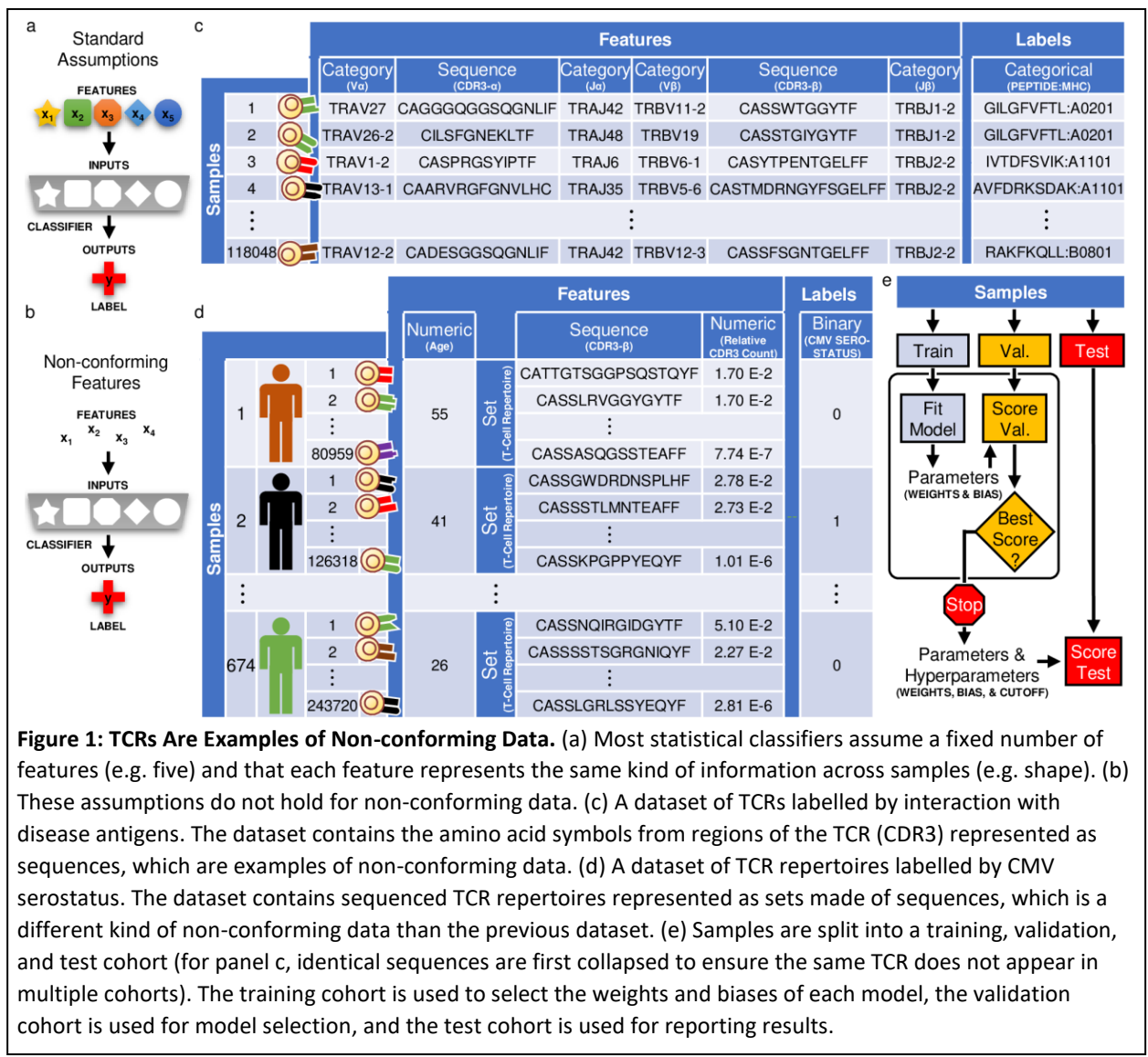

**Figure 1: TCRs Are Examples of Non-conforming Data.** (a) Most statistical classifiers assume a fixed number of features (e.g. five) and that each feature represents the same kind of information across samples (e.g. shape). (b) These assumptions do not hold for non-conforming data. (c) A dataset of TCRs labelled by interaction with disease antigens. The dataset contains the amino acid symbols from regions of the TCR (CDR3) represented as sequences, which are examples of non-conforming data. (d) A dataset of TCR repertoires labelled by CMV serostatus. The dataset contains sequenced TCR repertoires represented as sets made of sequences, which is a different kind of non-conforming data than the previous dataset. (e) Samples are split into a training, validation, and test cohort (for panel c, identical sequences are first collapsed to ensure the same TCR does not appear in multiple cohorts). The training cohort is used to select the weights and biases of each model, the validation cohort is used for model selection, and the test cohort is used for reporting results.

In this study, we adapt Iwana and Uchida's approach to handle a broader suite of non-conforming features, beginning with (mathematical) sets. The essential property of a set is that the content but not the order of the symbols conveys information, and therefore the order of the symbols has no meaning. Sets are examples of non-conforming data because there is no order indicating the position of each feature. By adapting their statistical classifier to also handle sets, our approach uses weights to both tune how each feature contributes to the prediction and place each feature into context in a manner invariant to order. We go a step further, combining approaches for sequences and sets into a composite model for sets made of sequences, which uses the weights to place each feature into context based on its order in the sequence while remaining invariant to the order of the sequences in each set. We call our approach *dynamic kernel matching* (DKM), providing a unified approach for classifying sequences, sets, sets of sequences, and potentially other non-conforming data as well. DKM also represents an extension of our previous approach for classifying sets, which we used to discover various disease signatures in patients' TCR sequences, by alleviating constraints that forced us to use only a single fixed-length subsequence from the top scoring TCR sequence to predict each label [4, 5, 6].

As examples of non-conforming data, we consider datasets of TCRs, anticipating these datasets to contain signatures for diagnosing disease. Other studies have avoided the problem of non-conforming features in TCR datasets by using handcrafted features, summary statistics, and preprocessing steps to reduce non-conforming data to conforming data, thereby producing conforming features that can be classified using standard methods [7 − 17]. While these strategies have been successful, there is no systematic approach that can guarantee essential information is not lost by reducing non-conforming data to conforming features. For this reason, approaches have been developed that directly classify non-conforming features rather than rely on a reduction step [1, 18 − 20]. Most of these approaches assume the non-conforming features follow a temporally directed relationship where the past contextualizes the present, but the present cannot contextualize the past. These assumptions may be inappropriate for datasets of TCRs, which do not necessarily contain notions of past and present. Because DKM does not impose notions of past and present, DKM may be ideal for datasets of TCRs. In this study, we showcase DKM on datasets of TCRs containing (i) sequences or (ii) sets of sequences, depending on what is labelled.

To demonstrate DKM's performance on classifying sequences, we fit (i) a multinomial regression model augmented with DKM to sequenced TCRs labelled by disease antigen (Fig. 1c) [21]. To demonstrate DKM's performance on classifying sets of sequences, we fit (ii) a logistic regression model augmented with DKM to sequenced TCR repertoires labelled by cytomegalovirus (CMV) serostatus (Fig. 1d) [15]. We then test the performance of our statistical classifier on samples from blindfolded cohorts (Fig. 1e). Finally, we analyze the statistical classifiers in the context of existing experimental data and discover these models may have identified biologically relevant patterns in the data. Our results collectively demonstrate DKM can be used to potentially uncover meaningful patterns in non-conforming features that predict the labels.

# Dynamic Kernel Matching

## *Representing Non-conforming Features*

As with any classification problem, the first step is to decide upon a way to represent the data, which in this study includes non-conforming features. We assume that each sample of non-conforming features consist of multiple symbols organized into some sort of structure, such as a sequence or a set.

We represent each symbol using a vector of $N$ numbers. For example, if we have a sequence of symbols, we replace each symbol by its vector representation, resulting in a sequence of vectors. Each time the same symbol appears, we use the same vector of numbers to represent that symbol, providing a consistent representation for each symbol across the dataset. Ideally, the numbers in each vector should describe properties of that symbol. In this study, the symbols represent amino acids and the numbers in each vector describe physicochemical properties of the amino acids, providing information necessary to determine, for example, that amino acid symbols R and K represent amino acids that both have positively charged sidechains interchangeable with respect to this common property (specifically, we use the five Atchley numbers to represent each amino acid [22]). We use $X$ to refer to the symbols in a sample, $i$ to act as an index of those symbols, $\vec{x}_i$ to refer to a vector representing a specific symbol in $X$ indexed by $i$, and $[x_{i,1}, x_{i,2}, \ldots, x_{i,N}]$ to represent the $N$ numbers in $\vec{x}_i$ (Supplementary Fig. 2a).

## *Representing Weights*

Suppose the statistical classifier is a linear regression model and therefore has weights for multiplying with features (DKM can also be used with a support-vector machine or deep neural network).

Because we are dealing with non-conforming features, the number of features is irregular and need not correspond to the number of weights. Therefore, we can pick the number of weights in our model irrespective of the number of features. We pick the number of weights to be evenly divisible by $N$ *(as defined in the previous section)*, allowing us to group the weights into vectors of $N$ weights. We use $\Theta$ to refer to these vectors, $j$ to act as an index of these vectors, $\vec{\theta}_j$ to refer to a specific vector in $\Theta$ indexed by $j$, and $[w_{j,1}, w_{j,2}, \ldots, w_{j,N}]$ to represent the $N$ weights contained by a vector $\vec{\theta}_j$ (Supplementary Fig. 2b).

By grouping the weights into vectors, we form symbols represented by $\vec{\theta}_j$. We will think of $\vec{\theta}_j$ as representing a symbol in $\Theta$, just as $\vec{x}_i$ represents a symbol in $X$. If we can identify symbols $\vec{x}_i$ to match to symbols $\vec{\theta}_j$ then we can assign the features in $\vec{x}_i$ to the weights in $\vec{\theta}_j$ because $\vec{x}_i$ and $\vec{\theta}_j$ are the same size $N$.

## *Similarity Score*

To decide which symbols in $X$ to match to symbols in $\Theta$, we need a *similarity score* to measure the similarity between these symbols. We define the similarity score between a symbol represented by $\vec{x}_i$ and a symbol represented by $\vec{\theta}_j$ as the sum of the features in $\vec{x}_i$ multiplied by the weights in $\vec{\theta}_j$.

$$S(\vec{x}_i, \vec{\theta}_j) = \sum_{k=1}^{N} x_{i,k} \cdot w_{j,k} \tag{1}$$

Using equation (1) to calculate each similarity score, we will match symbols in $X$ to symbols in $\Theta$ to maximize the overall similarity scores. Because each similarity score depends on weights, the weights influence how symbols in $X$ are matched to symbols in $\Theta$. As we will see, the algorithm for maximizing the overall similarity scores is different for sequences than sets, but after that the subsequent steps are the same regardless of which algorithm is used for each datatype.

### *Maximizing the Sum of Similarity Scores for Sequences*

Suppose that the symbols in X represent a sequence. In this case, we represent the symbols in $\Theta$ as a sequence so that we can use a *sequence alignment algorithm* to match symbols in X to symbols in $\Theta$. Sequence alignment algorithms pad sequences to place similar symbols from two sequences into matching positions. Similarity for each possible match is determined using a similarity score, like defined in equation (1). The sum of the similarity scores across every position is called the *alignment score*, and the objective of a sequence alignment algorithm is to find how to match symbols in the sequences to maximize this alignment score. Many sequence alignment algorithms have been developed, such as the Needleman–Wunsch algorithm used in this study, that guarantee finding an alignment with the maximum possible alignment score [23].

### *Maximizing the Sum of Similarity Scores for Sets*

Suppose that the symbols in X represent a set. In this case, we represent the symbols in $\Theta$ as a set so that we can solve the *assignment problem*, matching symbols in X to symbols in $\Theta$. We can think of solving the assignment problem as running an alignment algorithm, like before, except now the order of the symbols does not matter. Similarity for each possible match is determined using a similarity score, like defined in equation (1). We refer to the sum of the similarity scores between matched symbols as an alignment score, and the objective of the assignment problem is to maximize this alignment score. Algorithms exist for solving assignment problems, like the Hungarian method [24].

### *Maximizing the Sum of Scores for Sets Made of Sequences*

Suppose that the symbols in X represent a set made of sequences. In this case, we represent the symbols in $\Theta$ as a set made of sequences and combine the two previous algorithms to match symbols in X to symbols in $\Theta$. First, we match symbols from each sequence in X to symbols from each sequence in $\Theta$ using a sequence alignment algorithm. The similarity score between symbols is defined using equation (1). The sum of the similarity scores produced by aligning sequences is treated as a *sub*-alignment score. The sub-alignment scores serve as the similarity scores in the assignment problem, allowing us to match sequences. We refer to the sum of the similarity scores in the assignment problem as the alignment score, and the objective of the assignment problem is the same as maximizing this alignment score.

### *Generating Predictions*

Many statistical classifiers generate a prediction by calculating the sum of the features multiplied by the weights. We will see how we can reuse the alignment score from any of the previous algorithms to arrive at a similar calculation.

Because the alignment score is the sum of similarity scores over the matched symbols, we need to introduce sub-indices to indicate matched symbols. For example, if $\vec{x}_1$ is unmatched but $\vec{x}_2$ is matched to $\vec{\theta}_3$ then we use sub-indices $i_1 = 2$ and $j_1 = 3$ to represent the 1st match. If $\vec{x}_3$ is matched to $\vec{\theta}_4$ then we use sub-indices $i_2 = 3$ and $j_2 = 4$ to represent the 2nd match. Using sub-indices and letting $L$ represent the number of matches, we can represent the matched symbols.

$$
\begin{array}{cccc}
\vec{x}_{i_1} & \vec{x}_{i_2} & \cdots & \vec{x}_{i_L} \\
| & | & & | \\
\vec{\theta}_{j_1} & \vec{\theta}_{j_2} & \cdots & \vec{\theta}_{j_L}
\end{array}
$$

We can now write the alignment score as the sum of the similarity scores over the matched symbols.

$$
A = \sum_{l=1}^{L} S(\vec{x}_{i_l}, \vec{\theta}_{j_l}) = \sum_{l=1}^{L} \sum_{k=1}^{N} x_{i_l,k} \cdot w_{j_l,k} = \sum_{l=1,k=1}^{L,N} x_{i_l,k} \cdot w_{j_l,k} \tag{2}
$$

$A$ denotes the alignment score. On the left side of equation (2), the alignment score is written as the sum of similarity scores $S(\vec{x}_{i_l}, \vec{\theta}_{j_l})$ over the matched symbols. Each similarity score is written as a sum of the features multiplied by the weights using equation (1), resulting in a double summation. The double summation is flattened into a single summation on the right side of equation (2), resulting in a summation of features multiplied by weights. Thus, the alignment score is a sum of features multiplied by weights, representing the same calculation required by a linear regression model to arrive at a prediction.

It is tempting to use the unnormalized value of the alignment score $A$ to classify $X$ because the right side of equation (2) is a sum of features multiplied by weights. However, the magnitude of $A$ can vary depending on the number of symbols in $X$, which can be undesirable depending on the classification problem. If we assume the variances of the similarity scores at each of the $L$ positions are roughly the same, then the variance of $A$ proportionally varies by $\geq L$. Therefore, we normalize $A$ by dividing by the lower bound of the standard deviation, $\sqrt{L}$, using this value to classify $X$. For example, if we use logistic regression for the binary classification of a sequence, the logit would be defined as $\text{logit} = A/\sqrt{L}$ and the predicted probability would be $P = 1/(1 + e^{-\text{logit}})$. Because we normalize away information about $L$, we include $L$ as a feature in the statistical classifier. We also include a bias term $b$ in each logit, which is a free parameter that sets the baseline probability when all the features have a value of zero.

Given a method to generate a prediction, we can define the objective function for the problem as would be done for any other statistical classifier. Gradient optimization techniques can be used to find values for the weights and bias term that maximize the probability of achieving the correct predictions for each sample in a training cohort [25]. *See "Supplementary Text, Dynamic Kernel Matching" for parameter fitting and other details.*

### Confidence Cutoffs

Biomarker studies have used rule-in and rule-out cutoffs to diagnose patients with and without a condition. Patients not captured by these cutoffs are given an indeterminate diagnosis indicating the need for additional observation or exams. By providing an indeterminate diagnosis on uncertain cases, only the patients that can be diagnosed with a high degree of confidence receive a diagnosis, resulting in a higher classification accuracy [26, 27]. When conducting a blindfolded study, the labels for uncaptured samples must remain blindfolded. The classification accuracy is calculated only from the unblindfolded samples captured by the rule-in and rule-out cutoffs. When reporting the classification accuracy, the number of samples captured by the cutoffs must be reported too. If the number of captured samples is not reported, all samples are assumed to be used in the calculation of the classification accuracy.

Rather than use rule-in and rule-out cutoffs, we introduce an entropy-based cutoff, which is not limited to just binary classification problems but can be applied to multinomial and regression problems as well. With this approach, entropy is used to measure the confidence of each prediction, as defined below.

$$H_j = -\sum_{i=1}^{M} p_j^{(i)} \cdot \ln p_j^{(i)}$$

$H_j$ represents the entropy associated with the prediction for the $j^{\text{th}}$ sample, $M$ represents the number of label categories, $i$ indicates a specific label category, and $p_j^{(i)}$ represents the probability assigned to each category by the statistical classifier. We define $H_{\text{cutoff}}$ as the cutoff for capturing samples. If $H_j \leq H_{\text{cutoff}}$ the sample is classified because the confidence is high. Otherwise, the sample is not captured by the cutoff because the confidence is low. In this study, we start with a value for $H_{\text{cutoff}}$ large enough to ensure all the samples are initially captured and decrease $H_{\text{cutoff}}$ in increments of $0.01$ until we find that the accuracy over captured samples is ≥95% on a validation cohort. We then apply the cutoff to capture samples on a blindfolded test cohort, unblind ourselves to the captured samples, and compute the accuracy.

## Datasets

Datasets of TCR sequences provide examples of non-conforming features for diagnosing disease. Classifying TCRs requires handling (i) sequences or (ii) sets of sequences, depending on what is labelled.

### Antigen Classification Dataset

Individual TCRs can be labelled by the disease-relevant molecules it can bind, called antigens, consisting of a peptide presented on a major histocompatibility complex (pMHC). Because each TCR gene is generated by random insertions and deletions in what is called complimentary determining regions 3 (CDR3s) located between stochastically paired V- and J-gene segments, each TCR cannot be adequately characterized by its V- and J-genes without the CDR3 sequences as well. To solve what we call the *antigen classification problem*, we need a statistical classifier that can handle the CDR3 sequences of each TCR. 10x Genomics has published a dataset of sequenced TCRs barcoded by a panel of pMHCs (arranged on a dextramer) (Fig. 1c) [21]. The goal is to predict the pMHC from the sequenced TCR, requiring an approach to handle the sequences of amino acid symbols in the CDR3s. Of the 44 pMHCs in the dataset, only the six pMHCs interacting with at least 500 unique receptors are used for the antigen classification problem, which are GILGFVFTL:A0201, KLGGALQAK:A0301, RLRAEAQVK:A0301, IVTDFSVIK:A1101, AVFDRKSDAK:A1101, and RAKFKQLL:B0801. Because this dataset is imbalanced, samples have been weighted to represent each of the six pMHCs equally (see supplementary text for further details).

### Repertoire Classification Dataset

In circumstances where individual TCRs are not labelled by disease antigen, we can still label TCR sequences by the patient's disease status. Patterns exist that predict this label because a patient's set of TCR sequences, referred to as the TCR repertoire, constantly adjust to the presence of disease antigens, and therefore contain traces of past and ongoing immune responses to the underlying diseases. To solve what we call the *repertoire classification problem*, where the patients are labelled but the individual TCRs are not, we need a statistical classifier that can handle the set of CDR3 sequences in the TCR repertoire. Adaptive Biotechnologies has published a dataset of TCR repertoires sequenced from peripheral blood and labelled by CMV serostatus (Fig. 1d) [15]. The goal is to predict each patient's CMV serostatus from their TCR repertoire, requiring an approach to handle the set of CDR3 sequences from each patient. Because this dataset is imbalanced, samples have been weighted to represent CMV+ and CMV- cases equally (see supplementary text for further details).

## Results

### Antigen Classification Problem

To demonstrate that a statistical classifier augmented with DKM can classify sequences, we modify a multinomial regression model (Supplementary Fig. 3) and fit it to the antigen classification dataset (Fig. 1c) [21]. The dataset has been balanced to represent each category equally. At every gradient optimization step, the fit (as measured by KL-divergence) to samples from the training cohort steadily improves from 2.71 to 0.684 bits, demonstrating that even a multinomial regression model augmented with DKM can fit complex data (to put these numbers in perspective, six balanced outcomes like the categories in this dataset have an entropy of $\log_2 6 \approx 2.58$ bits). Measuring the fit to samples from the validation cohort at each gradient optimization step reveals a steady improvement from 2.60 to 1.18 bits, demonstrating the statistical classifier generalizes to holdout samples not used to fit the weight and bias values (Fig. 2a). At the end of the study, we unblindfold ourselves to samples from the test cohort and measure the fit as 1.31 bits, consistent with results from the validation cohort (Fig. 2a).

As a control, we permute the features with respect to the labels, eliminating any relationship between the features and labels, and refit the statistical classifier to verify that it no longer generalizes to holdout samples. On permuted data, the fit to samples from the training cohort steadily improves from 2.65 to 1.18 bits, representing the statistical classifier's capacity to memorize labels when there is no relationship between the features and labels (Fig. 2a). The fit as measured to samples from the validation cohort worsens from 2.91 to 3.14 bits, worse than the 2.54 bits that would be considered significant, indicating no ability to generalize to holdout samples when the statistical classifier memorizes labels (Fig. 2a). The fit to the test cohort under permutation is 3.17 bits, consistent with results from the validation cohort under permutation (Fig. 2a).

Next, we determine the classification accuracy on samples from the test cohort (without confidence cutoffs). We calculate the classification accuracy as the average number of times the most probable prediction matches the true label, averaged over all samples. On the test cohort, the accuracy is 70.5% (see Supplementary Fig. 4 for the confusion matrix). This classification accuracy is better than those achieved by standard classifiers using conforming features handcrafted for TCR data and even deep learning methods considered state of the art for many problems (Supplementary Fig. 6) [12, 17]. Because there are six possible outcomes, the baseline accuracy achievable by chance compares to guessing the outcome of a six-sided die roll, explaining the 16.6% ≈ 1/6 classification accuracy our approach achieves on permuted data.

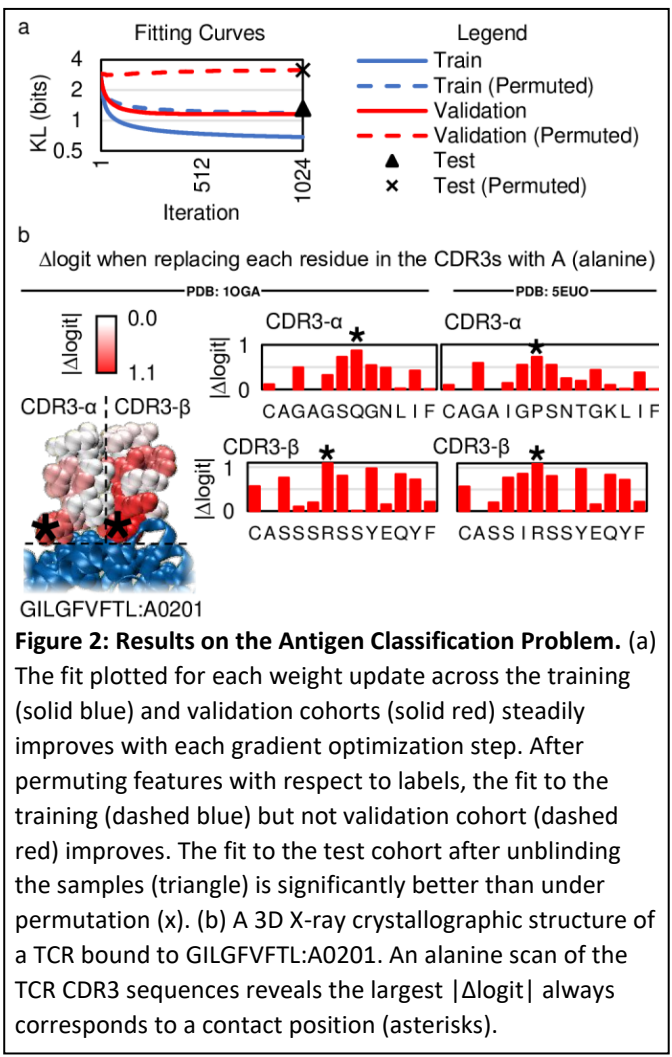

**Figure 2: Results on the Antigen Classification Problem.** (a) The fit plotted for each weight update across the training (solid blue) and validation cohorts (solid red) steadily improves with each gradient optimization step. After permuting features with respect to labels, the fit to the training (dashed blue) but not validation cohort (dashed red) improves. The fit to the test cohort after unblinding the samples (triangle) is significantly better than under permutation (x). (b) A 3D X-ray crystallographic structure of a TCR bound to GILGFVFTL:A0201. An alanine scan of the TCR CDR3 sequences reveals the largest |Δlogit| always corresponds to a contact position (asterisks).

We hypothesize the statistical classifier generates accurate predictions using specific non-conforming features describing molecular interactions between the TCR and pMHC. To attempt to answer this question, we search existing 3D X-ray crystallographic structures of TCR bound to any of the six pMHCs in the dataset, which would allow us to analyze the statistical classifier in the context of the observed molecular interactions between the TCR and pMHC. We find four unique TCRs bound to GILGFVFTL:A0201, a pMHC in our dataset [28, 29, 30]. The statistical classifier correctly predicts the four TCRs interact with this pMHC, but only two of the TCRs are captured by ≥95% confidence cutoffs (Supplementary Fig. 7a). We restrict our analysis to these two receptors, anticipating these cases to exhibit the clearest results. To ascertain the importance of each non-conforming feature, we conduct an in-silico alanine scan by systematically replacing each amino acid symbol in the CDR3s with alanine (symbol A), computing the change in logit, and then putting back the original amino acid symbol into the CDR3. The greatest logit changes tend to correspond to positions in the CDR3 in contact with pMHC (≥5Å), with the greatest logit change always corresponding to a contact position, as indicated by asterisks (Fig. 2b, Supplementary Fig. 7b,c). These observations agree with the hypothesis that the statistical classifier identifies and utilizes non-conforming features describing the molecular interactions between the TCR and pMHC.

### *Repertoire Classification Problem*

To demonstrate that a statistical classifier augmented with DKM can classify sets of made sequences, we modify a logistic regression model (Supplementary Fig. 8) and fit it to the repertoire classification dataset (Fig. 1d) [15]. The dataset has been balanced to represent each category equally. At every gradient optimization step, the fit (as measured by KL-divergence) to samples from the training cohort steadily improves from 0.966 to 0.728 bits, demonstrating a logistic regression model augmented with DKM can fit even this highly complex dataset (*we refit the model 128 times in an attempt to find the global optimum, using the best fit as measured on samples from the training cohort,* Supplementary Fig. 8j). Measuring the fit to samples from the validation cohort at each gradient optimization step reveals a steady improvement from 0.914 to 0.798 bits, demonstrating the statistical classifier generalizes to holdout samples not used to fit the weight and bias values (Fig. 3a). At the end of the study, we unblindfold ourselves to samples from the test cohort and measure the fit as 0.869 bits, consistent with results from the validation cohort (Fig. 3a).

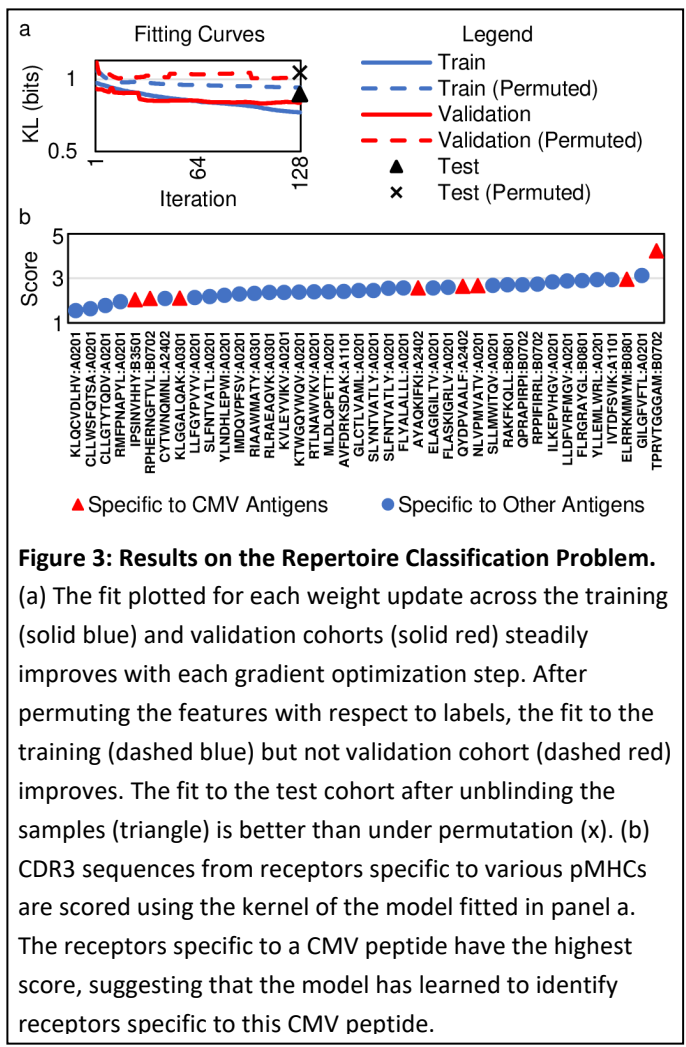

As a control, we permute the features with respect to the labels, eliminating any relationship between the features and labels, and refit the statistical classifier to verify that it no longer generalizes to holdout samples. On permuted data, the fit to samples from the training cohort steadily improves from 1.09 to 0.927 bits, representing the statistical classifier's capacity to memorize labels when there is no relationship between the features and labels (Fig. 3a). The fit as measured over samples from the validation cohort fluctuates between 1.18 to 1.016 bits, worse than the 1.0 bit that would be considered significant, indicating no ability to generalize to holdout samples when the statistical classifier memorizes labels (Fig. 3a). The fit on samples from the test cohort under permutation is 1.07 bits, consistent with results from the validation cohort under permutation (Fig. 3a).

**Figure 3: Results on the Repertoire Classification Problem.** (a) The fit plotted for each weight update across the training (solid blue) and validation cohorts (solid red) steadily improves with each gradient optimization step. After permuting the features with respect to labels, the fit to the training (dashed blue) but not validation cohort (dashed red) improves. The fit to the test cohort after unblinding the samples (triangle) is better than under permutation (x). (b) CDR3 sequences from receptors specific to various pMHCs are scored using the kernel of the model fitted in panel a. The receptors specific to a CMV peptide have the highest score, suggesting that the model has learned to identify receptors specific to this CMV peptide.

Next, we determine the classification accuracy on samples from the test cohort (without confidence cutoffs). We calculate the classification accuracy as the average number of times the most probable prediction matches the true label, balanced over the two categories (i.e. CMV+ and CMV-). On the test cohort, the accuracy is 67.6% (see Supplementary Fig. 9 for the ROC curve with AUC of 0.75). This classification accuracy is significantly worse than a previously reported result by Emerson and colleagues, where they developed an approach specific for TCR repertoires to predict the labels [15] (supplementary Fig. 10). Unfortunately, the source code to reproduce their results has not been made available for review.

We hypothesize the statistical classifier generates accurate predictions using specific non-conforming features describing TCRs that bind to CMV peptide. To answer this question, we use the weights and bias values to score all β-chain CDR3 sequences in the dataset published by 10x Genomics, allowing us to score and compare TCR CDR3 sequences interacting with CMV peptide to those interacting with non-CMV peptides (*essentially, the kernel of the statistical classifier fitted without individually labelled TCRs is being used to score TCRs interacting with known pMHC*). For missing features, like patient age, we use a value of zero. Because of these missing features, we can only evaluate the scores relative to other scores in the 10x Genomics dataset. For each pMHC in the 10x Genomics dataset, we average the scores for CDR3 sequences from TCRs that interact with the same pMHC. TPRVTGGGAM:B0702 has the largest score, which is a CMV peptide, congruent with our hypothesis that the statistical classifier identifies and utilizes non-conforming features describing TCRs that interact with CMV peptide (Fig. 3b). In fact, two of the top three scoring pMHCs contain CMV peptide. Other CMV peptides do not have comparably high scores, indicating the statistical classifier, fitted without using labels on individual TCRs, cannot identify receptors interacting with every possible CMV peptide.

### Applying Confidence Cutoffs

Even with a low classification accuracy, a statistical classifier can prove useful if it provides accurate positive and negative diagnoses for a subset of patients. With this goal in mind, we use confidence cutoffs to capture a subset of samples that can be classified with ≥95% accuracy. We find that cutoffs of $H_{cutoff}^{val} = 0.98$ and $H_{cutoff}^{val} = 0.527$ for the antigen and repertoire classification problems, respectively, achieve classification accuracies of ≥95% over samples captured from the validation cohort. Next, we run samples from the test cohort through the statistical classifiers, applying the confidence cutoffs $H_{cutoff}^{val}$ to capture samples from the test cohort. The labels are unblindfolded only for captured samples, and the classification accuracy is computed. On the antigen classification problem, the classification accuracy is 97.1% capturing 44.5% of test cohort samples, and on the repertoire classification problem, the classification accuracy is 96.0% capturing 18.0% of test cohort samples (Fig. 4).

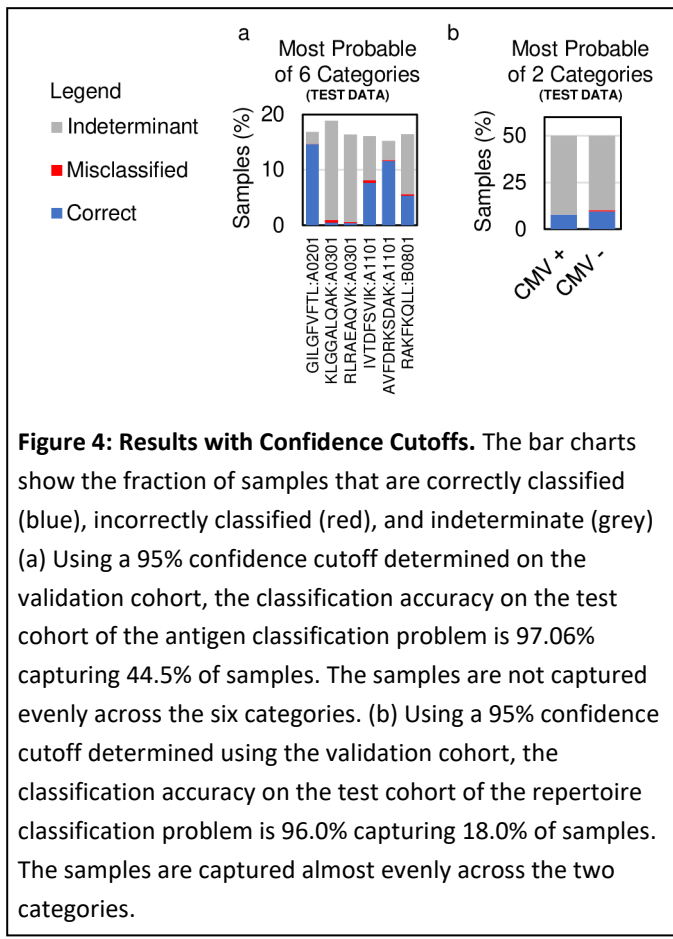

**Figure 4: Results with Confidence Cutoffs.** The bar charts show the fraction of samples that are correctly classified (blue), incorrectly classified (red), and indeterminate (grey) (a) Using a 95% confidence cutoff determined on the validation cohort, the classification accuracy on the test cohort of the antigen classification problem is 97.06% capturing 44.5% of samples. The samples are not captured evenly across the six categories. (b) Using a 95% confidence cutoff determined using the validation cohort, the classification accuracy on the test cohort of the repertoire classification problem is 96.0% capturing 18.0% of samples. The samples are captured almost evenly across the two categories.

We wondered if the confidence cutoffs capture only samples from a restricted subset of categories or if samples are captured evenly across all categories. On the antigen classification problem, captured samples are not balanced across categories (Fig. 4a). Nearly every sample for categories KLGGALQAK:A0301 and RLRAEAQVK:A0301 are left uncaptured. These two pMHCs represent the most challenging cases to distinguish given that the peptides are the same length sharing biochemically similar amino acid residues presented on the same MHC, explaining why the confidence in the predictions is not high enough to capture samples from these categories. On the repertoire classification dataset, the captured samples are approximately balanced across the two categories representing CMV+ and CMV- samples (Fig. 4b). Thus, whether captured samples evenly include all categories appears to depend on whether one category is more challenging to distinguish than others.

## Discussion

We demonstrate DKM can handle sequences, fitting a multinomial regression model augmented with DKM to sequenced TCRs labelled by interaction with disease antigen. The fitted model generalizes to samples in the test cohort, predicting the antigen with 70.5% accuracy, which is well above the 16.6% classification accuracy achievable by random chance and better than the alternative approaches we tried (Supplementary Fig. 4). To generate accurate predictions, we hypothesize the model must be using non-conforming features that describe molecular interactions between the TCR and pMHC. Using the sequences of TCRs in 3D X-ray crystallographic structures, we identify the amino acid residue in each CDR3 sequence most important to the correct prediction. We observe that this residue is always in contact with pMHC (≥5Å). These results are consistent with the hypothesis that DKM uses non-conforming features describing molecular interactions between the TCR and pMHC to generate accurate predictions.

Next, we demonstrate DKM can handle sets of sequences, fitting a logistic regression model augmented with DKM to sequenced TCR repertoires labelled by cytomegalovirus (CMV) serostatus. We essentially reuse the DKM component for the antigen classification problem, having demonstrated that it handles the CDR3 sequences, and combine it with an approach for handling sets, to create an approach to solve the repertoire classification problem. On a test cohort, the fitted model achieves a classification accuracy of only 67.6%. To address this low accuracy, we use confidence cutoffs to capture samples that can be accurately diagnosed, excluding predictions where the uncertainty is too high. Using confidence cutoffs, approximately 8% of patients receive a CMV+ diagnosis, 10% receive a CMV- diagnosis, and 82% receive an indeterminate diagnosis. For the 18% of patients that are diagnosed, the classification accuracy is 96%. To generate accurate predictions, we hypothesize the model must be identifying TCRs specific to CMV antigen. To test this hypothesis, we score CDR3 sequences from TCRs interacting with known pMHCs using the weights of the fitted model. We observe that TCRs interacting TPRVTGGGAM, a CMV peptide, receive the highest score. These results are consistent with the hypothesis that the model uses TCRs specific to at least this CMV antigen to generate accurate predictions.

We find it remarkable that DKM can be used to build models to classify both sequences and sets of sequences, suggesting DKM can be reused on other datasets irrespective of the kind of data. We envision DKM providing a unified approach for handling many other kinds of non-conforming data and look forward to collaborating with other researchers to help build statistical classifiers for their problems.

## Materials and Methods

Detailed instructions for downloading the data and building the datasets are included in the README of the source code accompanying this manuscript and available at http://github.com/jostmey/dkm. Briefly, data for the antigen classification problem can be downloaded directly from 10x Genomics webpage after creating a free account with 10x Genomics, and data for the repertoire classification problem can be downloaded from https://doi.org/10.21417/B7001Z after creating a free account with Adaptive Biotechnologies.

Downloaded sequencing data is used without imposing additional quality controls. The default threshold for the read counts provided in the 10x Genomics dataset are used (although the optimal value for this threshold is a matter of debate because the threshold effects the false and true positive rates as well as the number of TCRs assigned to each pMHC). Descriptions for shuffling and splitting the samples into training, validation, and test cohorts is available in the supplementary text accompanying this manuscript and can be inferred from the source code.

The statistical classifiers in this study are implemented in TensorFlow v1.14 and run on a pair of RTX 2080 Ti GPUs or a cluster of either 8 P100 or V100 GPUs. Source code for the statistical classifiers accompanies this manuscript and is available at http://github.com/jostmey/dkm.

The 3D X-ray crystallographic structures analyzed in this study can be found at https://www.rcsb.org/ using the protein databank identifiers 1OGA, 5EUO, 5E6I, and 5ISZ. Renderings of the protein structures depicted in the figures are generated using Visual Molecular Dynamics (VMD).

## Acknowledgements

We are grateful that the antigen and repertoire classification datasets have been made available online by 10x Genomics and Adaptive Biotechnologies. Computing time on the UT Southwestern BioHPC computing cluster is made available through the Harold C. Simmons Comprehensive Cancer Center. This work is supported by the Department of Population and Data Sciences at UT Southwestern.

## References

[1] H. Shimodaira, K.-i. Noma, M. Nakai and S. Sagayama, "Dynamic Time-Alignment Kernel in Support Vector Machine," in *Advances in Neural Information Processing Systems 14*, 2001.

[2] B. K. Iwana, V. Frinken and S. Uchida, "A Robust Dissimilarity-Based Neural Network for Temporal Pattern Recognition," in *2016 15th International Conference on Frontiers in Handwriting Recognition (ICFHR)*, 2016.

[3] B. K. Iwana, V. Frinken and S. Uchida, "DTW-NN: A novel neural network for time series recognition using dynamic alignment between inputs and weights," *Knowledge Based Systems,* vol. 188, p. 104971, 2020.

[4] J. Ostmeyer, S. Christley, W. H. Rounds, I. Toby, B. M. Greenberg, N. L. Monson and L. G. Cowell, "Statistical classifiers for diagnosing disease from immune repertoires: a case study using multiple sclerosis.," *BMC Bioinformatics,* vol. 18, no. 1, p. 401, 2017.

[5] J. Ostmeyer, S. Christley, I. T. Toby and L. G. Cowell, "Biophysicochemical motifs in T-cell receptor sequences distinguish repertoires from tumor-infiltrating lymphocyte and adjacent healthy tissue," *Cancer Research,* vol. 79, no. 7, pp. 1671-1680, 2019.

[6] J. L. Ostmeyer, L. G. Cowell and S. Christley, "Developing and validating an approach for diagnosing and prognosticating cancer from biochemical motifs in T-cell receptors.," *Journal of Clinical Oncology,* vol. 38, 2020.

[7] E. M. Cameron, S. Spencer, J. Lazarini, C. T. Harp, E. S. Ward, M. Burgoon, G. P. Owens, M. K. Racke, J. L. Bennett, E. M. Frohman and N. L. Monson, "Potential of a unique antibody gene signature to predict conversion to clinically definite multiple sclerosis," *Journal of Neuroimmunology,* vol. 213, no. 1, pp. 123-130, 2009.

[8] M. Cinelli, Y. Sun, K. Best, J. M. Heather, S. Reich-Zeliger, E. Shifrut, N. Friedman, J. Shawe-Taylor and B. Chain, "Feature selection using a one dimensional naïve Bayes' classifier increases the accuracy of support vector machine classification of CDR3 repertoires," *Bioinformatics,* vol. 33, no. 7, pp. 951-955, 2017.

[9] V. Greiff, P. Bhat, S. C. Cook, U. Menzel, W. Kang and S. T. Reddy, "A bioinformatic framework for immune repertoire diversity profiling enables detection of immunological status," *Genome Medicine,* vol. 7, no. 1, pp. 49-52, 2015.

[10] V. Greiff, C. R. Weber, J. Palme, U. Bodenhofer, E. Miho, U. Menzel and S. T. Reddy, "Learning the High-Dimensional Immunogenomic Features That Predict Public and Private Antibody Repertoires," *Journal of Immunology,* vol. 199, no. 8, pp. 2985-2997, 2017.

[11] H. M. Li, T. Hiroi, Y. Zhang, A. Shi, G. Chen, S. De, E. Metter, W. H. Wood, A. Sharov, J. D. Milner, K. G. Becker, M. Zhan and N. P. Weng, "TCRβ repertoire of CD4+ and CD8+ T cells is distinct in

richness, distribution, and CDR3 amino acid composition," *Journal of Leukocyte Biology,* vol. 99, no. 3, pp. 505-513, 2016.

[12] N. D. Neuter, W. Bittremieux, C. Beirnaert, B. Cuypers, A. Mrzic, P. Moris, A. Suls, V. V. Tendeloo, B. Ogunjimi, K. Laukens and P. Meysman, "On the feasibility of mining CD8+ T cell receptor patterns underlying immunogenic peptide recognition," *research in computational molecular biology,* vol. 70, no. 3, pp. 159-168, 2018.

[13] B. J. Olson, P. Moghimi, C. Schramm, A. Obraztsova, D. K. Ralph, J. A. V. Heiden, M. Shugay, A. J. Shepherd, W. D. Lees and F. A. M. Iv, "sumrep: a summary statistic framework for immune receptor repertoire comparison and model validation," *Frontiers in Immunology,* vol. 10, p. 2533, 2019.

[14] N. Thomas, K. Best, M. Cinelli, S. Reich-Zeliger, H. Gal, E. Shifrut, A. Madi, N. Friedman, J. Shawe-Taylor and B. Chain, "Tracking global changes induced in the CD4 T cell receptor repertoire by immunization with a complex antigen using short stretches of CDR3 protein sequence," *Bioinformatics,* vol. 30, no. 22, pp. 3181-3188, 2014.

[15] R. O. Emerson, W. S. DeWitt, M. Vignali, J. Gravley, J. K. Hu, E. J. Osborne, C. Desmarais, M. Klinger, C. S. Carlson, J. A. Hansen, M. Rieder and H. S. Robins, "Immunosequencing identifies signatures of cytomegalovirus exposure history and HLA-mediated effects on the T cell repertoire," *Nature Genetics,* vol. 49, no. 5, pp. 659-665, 2017.

[16] H. Konishi, D. Komura, H. Katoh, S. Atsumi, H. Koda, A. Yamamoto, Y. Seto, M. Fukayama, R. Yamaguchi, S. Imoto and S. Ishikawa, "Capturing the differences between humoral immunity in the normal and tumor environments from repertoire-seq of B-cell receptors using supervised machine learning," *BMC Bioinformatics,* vol. 20, no. 1, p. 267, 2019.

[17] P. Dash, A. J. Fiore-Gartland, T. Hertz, G. C.-C. Wang, X. Sharma, A. Souquette, J. C. Crawford, E. B. Clemens, T. H. Nguyen, K. Kedzierska, N. L. L. Gruta, P. Bradley and P. G. Thomas, "Quantifiable predictive features define epitope-specific T cell receptor repertoires," *Nature,* vol. 547, no. 7661, pp. 89-93, 2017.

[18] S. Hochreiter and J. Schmidhuber, "Long short-term memory," *Neural Computation,* vol. 9, no. 8, pp. 1735-1780, 1997.

[19] N. Kalchbrenner, E. Grefenstette and P. Blunsom, "A Convolutional Neural Network for Modelling Sentences," in *Proceedings of the 52nd Annual Meeting of the Association for Computational Linguistics (Volume 1: Long Papers)*, 2014.

[20] H. Zeng, M. D. Edwards, G. Liu and D. K. Gifford, "Convolutional neural network architectures for predicting DNA–protein binding," *Bioinformatics,* vol. 32, no. 12, pp. 121-127, 2016.

[21] 10xGenomics, A New Way of Exploring Immunity - Linking Highly Multiplexed Antigen Recognition to Immune Repertoire and Phenotype.

[22] W. R. Atchley, J. Zhao, A. D. Fernandes and T. Drüke, "Solving the protein sequence metric problem," *Proceedings of the National Academy of Sciences of the United States of America,* vol. 102, no. 18, pp. 6395-6400, 2005.

[23] S. B. Needleman and C. D. Wunsch, "A general method applicable to the search for similarities in the amino acid sequence of two proteins," *Journal of Molecular Biology,* vol. 48, no. 3, pp. 443-453, 1970.

[24] H. W. Kuhn, "The Hungarian method for the assignment problem," *Naval Research Logistics Quarterly,* vol. 2, no. 1, pp. 83-97, 1955.

[25] D. P. Kingma and J. L. Ba, "Adam: A Method for Stochastic Optimization," in *ICLR 2015 : International Conference on Learning Representations 2015*, 2015.

[26] A. R. Feinstein, "The inadequacy of binary models for the clinical reality of three-zone diagnostic decisions," *Journal of Clinical Epidemiology,* vol. 43, no. 1, pp. 109-113, 1990.

[27] C. Shortt, J. Ma, N. Clayton, J. Sherbino, R. Whitlock, G. Pare, S. A. Hill, M. McQueen, S. R. Mehta, P. Devereaux, A. Worster and P. A. Kavsak, "Rule-In and Rule-Out of Myocardial Infarction Using Cardiac Troponin and Glycemic Biomarkers in Patients with Symptoms Suggestive of Acute Coronary Syndrome," *Clinical Chemistry,* vol. 63, no. 1, pp. 403-414, 2017.

[28] G. Chen, X. Yang, A. Ko, X. Sun, M. Gao, Y. Zhang, A. Shi, R. A. Mariuzza and N.-p. Weng, "Sequence and Structural Analyses Reveal Distinct and Highly Diverse Human CD8+ TCR Repertoires to Immunodominant Viral Antigens," *Cell Reports,* vol. 19, no. 3, pp. 569-583, 2017.

[29] I. Y. Song, A. Gil, R. Mishra, D. Ghersi, L. K. Selin and L. J. Stern, "Broad TCR repertoire and diverse structural solutions for recognition of an immunodominant CD8 + T cell epitope," *Nature Structural & Molecular Biology,* vol. 24, no. 4, pp. 395-406, 2017.

[30] G. Stewart-Jones, A. Mcmichael, J. Bell, D. Stuart and E. Jones, "A Structural Basis for Immunodominant Human T Cell Receptor Recognition," *Nature Immunology,* vol. 4, no. 7, pp. 657-663, 2003.

## Supporting Information

| | Weights Features [a] | Order Invariant [b] | Sensitive to Order [c] | Undirected [d] |
|---|---|---|---|---|
| Bag of Words Model | Weights Words | ✔ | ✘ | ✔ |
| Multiple Instance Learning | Model Dependent | ✔ | ✘ | ✔ |
| Nearest Neighbor (BY SEQUENCE ALIGNMENT SCORE) | ✘ | ✘ | ✔ | ✔ |
| Hidden Markov Models | Transitions Weighted | ✘ | ✔ | ✘ |
| Recurrent Neural Networks (e.g. LSTM) | ✔ | ✘ | ✔ | ✘ |
| Convolutional Neural Networks (WITH GLOBAL POOLING) | ✔ | Globally Invariant [e] | Locally Sensitive [e] | ✔ |
| Dynamic Kernel Matching (SEQUENCE ALIGNMENT) | ✔ | ✘ | ✔ | ✔ |
| Dynamic Kernel Matching (SET ALIGNMENT) | ✔ | ✔ | ✘ | ✔ |

**Supplementary Figure 1:** Figure summarizing some major properties of statistical classifiers for variable-length data. (a) Statistical classifiers that use weights can tune the contribution of each feature in the prediction, minimizing the contribution of features that lead to misclassifications. (b) Statistical classifiers that remain invariant to the order of the features are suitable for classifying sets. (c) Statistical classifiers that are sensitive to the order of the features are suitable for classifying sequences (assuming variable-length data). (d) Undirected statistical classifiers achieve the same performance after refitting the model when the order of the features is reversed in every sample (For example, we would not want the performance to drop because we refit the model after flipping every CDR3 sequence in a dataset). Recurrent neural networks (RNNs) are an example of models that are not undirected. The feedforward connections of RNNs impose directionality on the way sequences or other data are read. (e) Global pooling has been used with convolutional neural networks (CNNs) to deal with variable length data, like sequences and sets. Convolutional filters are sensitive to the order of the features spanned by the width of the filter. However, the activations of the filter across each sequence are pooled globally, discarding information about the order of activation patterns along each sequence. Thus, CNNs are sensitive to local patterns spanned by its filters but not global patterns in the activations of those filters.

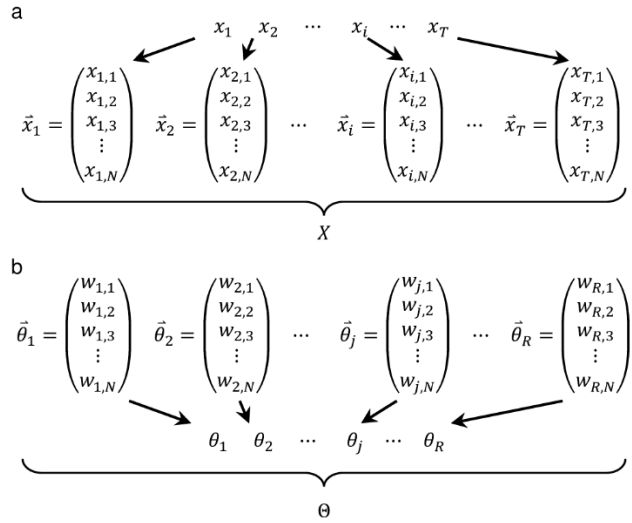

**Supplementary Figure 2:** Representation of the variables for the non-conforming features and weights of a DKM model. (a) Let us assume the non-conforming features consists of a sequence (or set) of $T$ symbols $x_1, x_2, \ldots x_i, \ldots x_T$. We replace each symbol with a vector of $N$ numbers describing that symbol, resulting in a sequence (or set) of $T$ vectors $\vec{x}_1, \vec{x}_2, \ldots \vec{x}_i, \ldots \vec{x}_T$. We use $X$ to refer to a container holding these vectors. (b) Let us assume the number of weights for our statistical classifier can form $R$ groups of $N$ weights. Each group of $N$ weights is used to form a vector, allowing us to write the weights as the vectors $\vec{\theta}_1, \vec{\theta}_2, \ldots \vec{\theta}_j, \ldots \vec{\theta}_R$. We can think of each vector as a symbol, forming the symbols $\theta_1, \theta_2, \ldots \theta_j, \ldots \theta_R$. We use $\Theta$ to refer to a container holding these vectors.

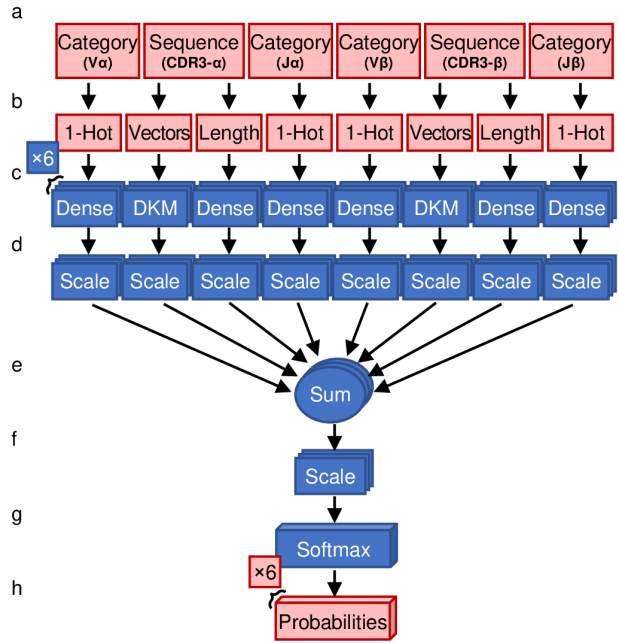

**Supplementary Figure 3:** Schematic representation of the multinomial regression model augmented with DKM for the antigen classification problem. (a) Features for each TCR are partitioned into six groups. (b) The representations for each feature group. (c) Dense modules assign a weight to each feature and compute a dot product. The DKM modules use a sequence alignment algorithm to match features with weights and compute a dot product. (d) Each dot product is normalized over training samples. The layout is repeated six times, one for each category of label. (e) The scaled dot products are added together into one dot product. (f) The resulting dot product is normalized over training samples. (g) The dot product is part of a multinomial logit and passed through a softmax function. (h) The softmax function produces probabilities representing the predictions of the statistical classifier. The six probabilities, one for each category, always sum to one.

Confusion Matrix

**Supplementary Figure 4:** Confusion matrix for visualizing the performance of the multinomial regression model augmented with DKM on samples from the test cohort. The number in each cell indicates the fraction of samples where the output of the model agrees with the true label. A value of 1 would indicate complete agreement and a value of 0 would indicate no agreement. The statistical classifier does well predicting labels for GILGFVFTL:A0201, IVTDFSVIK:A1101, AVFDRKSDAK:A1101, and RAKFKQLL:B0801. Most of the misclassified samples are labelled either KLGGALQAK:A0301 or RLRAEAQVK:A0301, which represent the most challenging cases to distinguish given that the peptides are the same length sharing biochemically similar amino acid residues presented on the same MHC. The statistical classifier correctly distinguishes between IVTDFSVIK:A1101 and AVFDRKSDAK:A1101, which is significant because the peptides in these two cases are presented on the same MHC, indicating the statistical classifier predicts the peptide, not the MHC type.

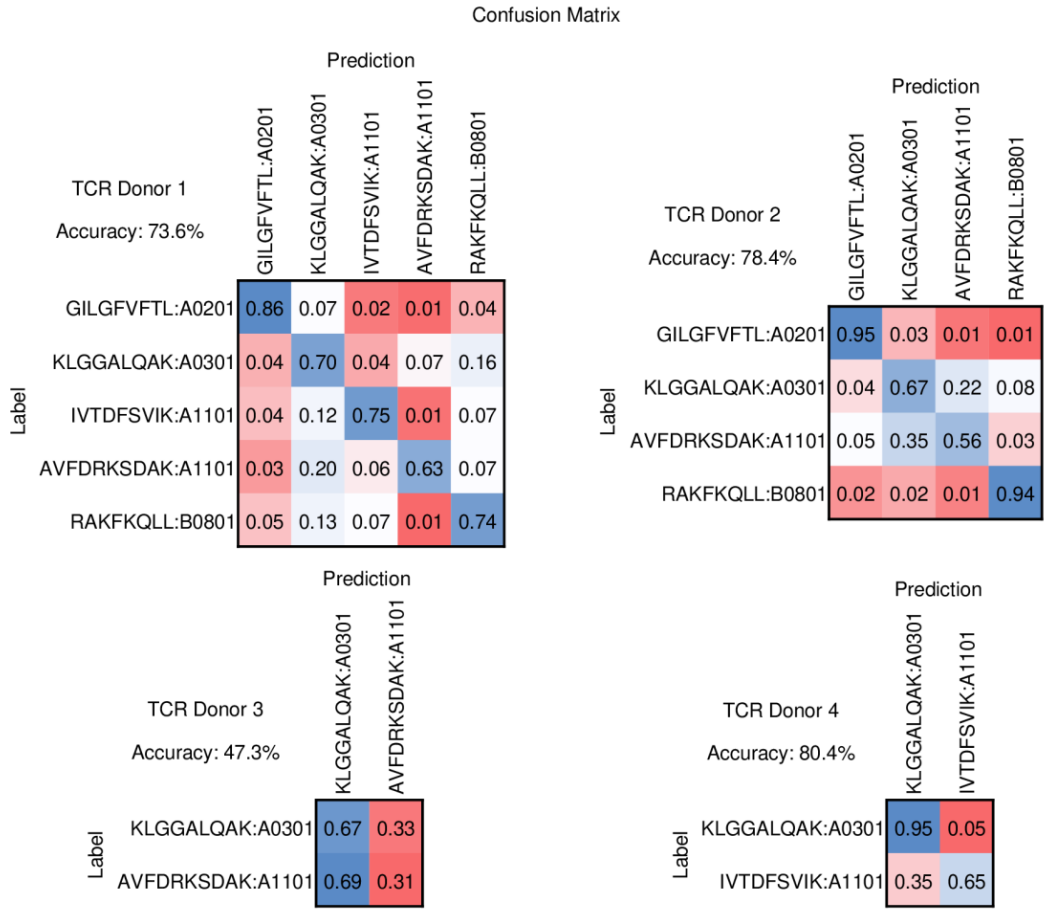

**Supplementary Figure 5:** Confusion matrices for visualizing the performance of the multinomial regression model augmented with DKM, where the TCRs are separated by the four TCR donors before training and evaluating the DKM model (pMHCs are from the dextramer constructs, not the TCR donors). Separating TCRs by the four TCR donors removes any possibility that the model learns to predict the donor rather than interaction with pMHC (some donors have greater numbers of TCRs interacting with specific pMHCs, resulting in a potential confounding factor). Separating TCRs by TCR donor reduced the number of TCRs associated with each pMHC, forcing many categories to fall below our threshold of 500 unique TCRs for inclusion as a category. For this reason, some of the original six categories do not appear with all four TCR donors. When restricted to TCRs from a single TCR donor, the DKM model does well on TCR donors 1, 2, and 4. On TCR donor 3, the DKM model fails to distinguish between KLGGALQAK:A0301 and RLRAEAQVK:A0301, which represent the most challenging cases to distinguish given that the peptides are the same length sharing biochemically similar amino acid residues presented on the same MHC.

Antigen Classification Problem

| | Model | KL-Divergence (bits) | Accuracy (%) |
|---|---|---|---|
| NN | Distance Metric from Pradyot Dash et. al. (NATURE 2017) | Not Applicable | 67.87 |
| RF | Handcrafted Features from De Neuter et. al. (IMMUNOGENETICS 2017) | Not Applicable | 67.35 |
| MR | * Handcrafted features applied on α-chain in addition to the β-chain | 1.447 | 65.51 |
| DNN | | 1.428 | 67.12 |
| RNN | LSTM | 1.395 | 60.77 |
| | GRU | 1.444 | 65.49 |
| CNN | Conv(size=4) – Global Pool | 1.533 | 62.14 |
| | Conv(size=4) – Conv(size=4) – Global Pool | 1.548 | 65.48 |
| DKM | DKM (size=32) * | 1.311 | 70.49 |
| | DKM (kmer=5, size=28) | 1.300 | 71.56 |

**Supplementary Figure 6:** Comparison of statistical classifiers on the antigen classification problem. All models are fitted to the training cohort and results reported on the test cohort. The considered models are NN (nearest neighbors), RF (random forest), MR (multinomial regression), DNN (deep neural network), RNN (recurrent neural network), CNN (convolutional neural network), and DKM (dynamic kernel matching). DKM models have the best fit over the test cohort and the highest classification accuracies. The DKM model with the asterisk is the model reported in the main text. The other DKM model classifies each CDR3 as a sequence of 5-mers.

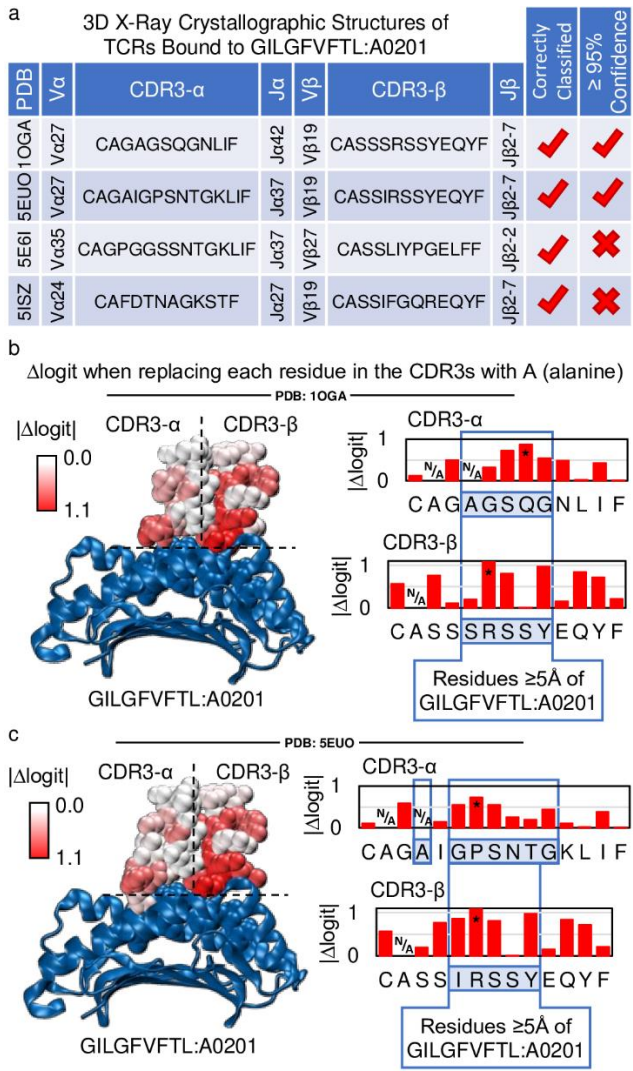

**Supplementary Figure 7:** (a) Four 3D X-ray crystallographic structures found with a TCR bound to GILGFVFTL:A0201. All four TCRs are correctly classified and two with ≥95% confidence (see *Applying Confidence Cutoffs*). (b) An alanine scan of the 3D X-ray crystallographic structure 1OGA (four letter codes are the protein databank identifier at https://www.rcsb.org/) reveals the |Δlogit| tend to be greater (redder) for pMHC (green) contact positions (≤5Å) than non-contact positions. The largest |Δlogit| for each CDR3 is always a contact position. (c) Same as before for another 3D X-ray crystallographic structure 5EUO.

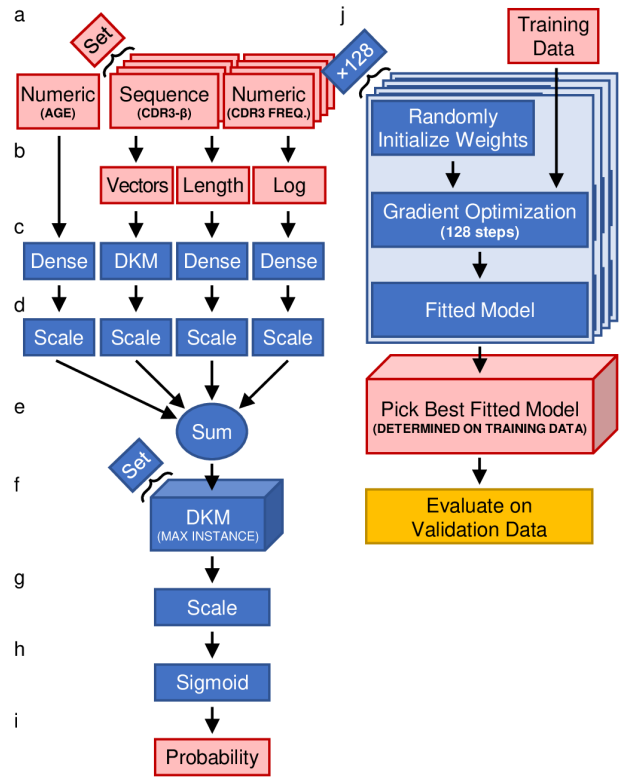

**Supplementary Figure 8:** Schematic representation of the logistic regression model augmented with DKM for the repertoire classification problem. (a) Features for each patient are partitioned into three groups. The last two feature groups, representing the CDR3-β sequence and relative CDR3 frequency, form a set. Each member of the set is passed through the model. (b) The representations for each feature group. (c) Dense modules assign a weight to each feature and compute a dot product. The DKM modules use a sequence alignment algorithm to match features with weights and compute a dot product. (d) Each dot product is normalized over training samples. (e) The scaled dot products are added together into one dot product. (f) This DKM module takes the maximum value over the set, which is the computation required to solve the assignment problem for the special case where Θ contains just one sequence (g) The maximum value is normalized over training samples. (h) The scaled value is passed through a sigmoid function. (i) The sigmoid function produces a probability representing the prediction of the statistical classifier. (j) Random values for each weight are refined by 128 steps of gradient optimization to produce a fitted model. Weights from the best of 128 attempts to fit the model are used to evaluate the model on the validation cohort and eventually the test cohort.

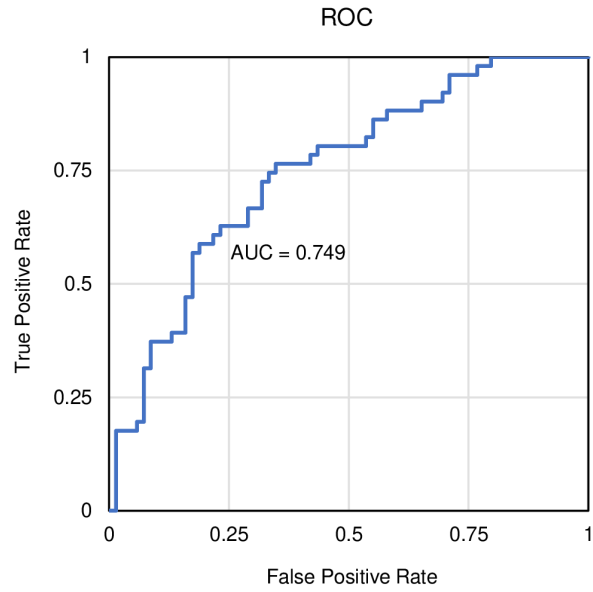

**Supplementary Figure 9:** Receiver Operating Characteristic (ROC) curve computed from a logistic regression model augmented with DKM on the test samples from the repertoire classification problem. The ROC curve shows true and false positive rates for different diagnostic thresholds. The area under the curve is 0.749.

Repertoire Classification Problem

| Model | Log-Likelihood (bits) | Accuracy (%) | Captured (%) | Accuracy (%) |
|---|---|---|---|---|
| | | | ≥ 95% Confidence Cutoff | |
| Emerson et al. **(NATURE GENETICS, 2017)** | Not Reported | 89% | Source code unavailable | |
| Max Snippet Model **(MAX SCORING 4-MER)** | -0.878 | 68.3% | Failed to meet cutoff | |
| DKM * **(SIZE = 1 SEQUENCE, 8 STEPS)** | -0.869 | 67.6% | 18% | 96% |
| DKM **(SIZE = 1 SEQUENCE, 32 STEPS)** | -0.831 | 73.5% | Failed to meet cutoff | |
| DKM **(SIZE = 2 SEQUENCES, 8 STEPS)** | -0.819 | 73.5% | Failed to meet cutoff | |

**Supplementary Figure 10:** Comparison of applicable statistical classifiers on the repertoire classification problem. All models are fitted to the training cohort, confidence cutoffs identified using the validation cohort, and results reported on the test cohort. The approach published by Emerson et al. achieves the best classification accuracy. We could not calculate the log-likelihood fit or apply confidence cutoffs because the source code to their method is unavailable. We use the classification accuracy as reported in their publication. The Max Snippet Model, as described in our earlier publications, has the worst performance [1, 2, 3]. No matter how stringent we make the confidence cutoffs, we cannot achieve a classification accuracy of ≥95% over captured samples. The asterisk indicates the model reported in the main text. Using confidence cutoffs, we achieve a classification accuracy of 96%, capturing 18% of patients. The performance is also shown for a DKM model where the weights are arranged into 32 steps instead of 8, and for another DKM model where the weights are arranged into two sequences instead of one. Both these DKM models achieve better performances on the test cohort than the DKM model we report in the main text. For these two variations of the DKM model, we could only achieve a classification accuracy of ≥90% using confidence cutoffs.

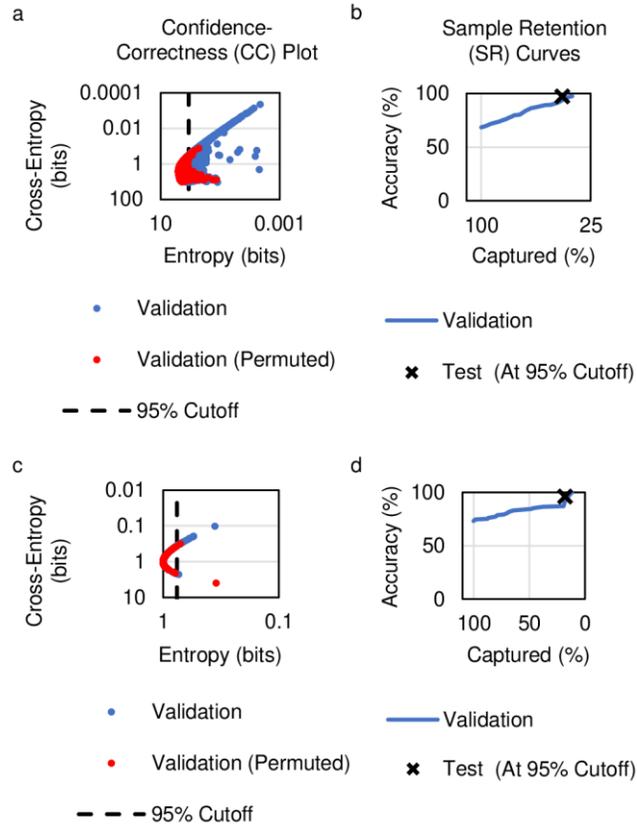

**Supplementary Figure 11:** (a) Using entropy as a measure of confidence, which is computed without labels, and cross entropy as a measure of correctness, which is computed with labels, we see a correlation between the confidence and correctness over each prediction (blue dots) of the statistical classifier on the antigen classification problem. Because the correlation exists, we can use the confidence to enrich for correctness, where the former is computed without knowing the labels. Samples captured to the right of the dashed line are classified with >95% accuracy (Fig. 4), while samples to the left are considered indeterminant. No correlation exists for permuted data (red dots). (b) Sample retention curves show the classification accuracy as the confidence cutoff is increased, reducing the number of captured samples. Using the 95% cutoff computed using the validation cohort, samples from the test cohort are captured and the accuracy computed ("x"). (c) Like panel a, for the repertoire classification problem. (d) Like panel b, for the repertoire classification problem.

## Supplementary Text, Dynamic Kernel Matching

### *Maximizing the Sum of Scores for Graphs*

Suppose the symbols in X represent a graph. In this case, we represent the symbols in Θ as a graph so that we can use a *graph alignment algorithm* to match symbols in X to symbols in Θ. We can think of the graph alignment algorithm as like running a sequence alignment algorithm. Similarity for each possible match is determined using a similarity score, like defined in equation (1). We refer to the sum of the similarity scores between matched symbols as an alignment score, and the objective of the graph alignment algorithm is to maximize this alignment score. Unfortunately, the general problem of graph alignment is regarded as NP-hard, which is why it is important to consider special cases where symbols can be matched together in a computationally efficient manner. For example, sequences and sets are examples of *path* and *bipartite* graphs, representing special cases where the symbols can be efficiently matched in polynomial time.

### *Discouraging Unmatched Symbols*

Algorithms for matching symbols may leave some symbols unmatched, which are represented as gaps by sequence alignment algorithms. To discourage unmatched symbols, we can introduce penalties incurred by the objective function of these algorithms whenever a symbol is left unmatched. For example, when $X$ is shorter than Θ and we want to ensure every symbol in $X$ is used, we can introduce a penalty $G_x$ approximating $-\infty$, constraining the alignment algorithm to use every symbol in $X$ to avoid the penalty. Alternatively, when Θ is shorter than $X$ and we want to ensure every symbol in Θ is used, we can introduce a penalty $G_\theta$ approximating $-\infty$, constraining the alignment algorithm to use every symbol in Θ to avoid the penalty.

### *Data Splits*

On the antigen classification problem, identical TCR sequences are collapsed before splitting the TCR sequences into a training, validation, and test cohort using a 60/20/20 split. On the repertoire classification problem, the patients are separated into Cohort I and Cohort II. Patients in Cohort I are randomly shuffled. 120 patients in Cohort I are set aside for the validation cohort while the remaining patients in Cohort I are used as the training cohort. Patients in Cohort II are used as the test cohort.

### *Data Balancing*

The dataset for the antigen classification problem is highly imbalanced because some pMHCs interact with significantly more TCRs than others. Sample frequencies $f_j$ assigned to each TCR balance the expected observation of each pMHC to $1/M_{\text{pMHC}}$, where $M_{\text{pMHC}}$ denotes the number of pMHCs.

$$f_j^{(i)} = \frac{c_j^{(i)}}{\sum_{j'} c_{j'}^{(i)}} \quad f_j = \frac{1}{M_{\text{pMHC}}} \sum_{i=1}^{M_{\text{pMHC}}} f_j^{(i)}$$

Here, $c_j^{(i)}$ is the number of cases where the $j^{\text{th}}$ TCR interacts with the $i^{\text{th}}$ pMHC, $f_j^{(i)}$ is the frequency of the $j^{\text{th}}$ TCR relative to all other TCRs interacting with the $i^{\text{th}}$ pMHC, and $f_j$ is calculated as the average of $f_j^{(i)}$ across all pMHCs. Because a TCR can interact with multiple pMHCs, the observed probability of interacting with the $i^{\text{th}}$ pMHC is computed as $y_j^{(i)} = c_j^{(i)} / \sum_{i'=1}^{M_{\text{pMHC}}} c_j^{(i')}$, which serves as the labels for this dataset.

The dataset for the repertoire classification problem is also imbalanced because a slightly greater number of patients are CMV+. Sample frequencies $f_j$ assigned to each patient $j$ balance the expected observation of CMV+ to CMV-.

$$f_j = \frac{1}{2} \cdot \begin{cases} \dfrac{1}{M_{\text{CMV+}}} & \text{for CMV+ patients} \\[2mm] \dfrac{1}{M_{\text{CMV-}}} & \text{for CMV- patients} \end{cases}$$

$M_{\text{CMV+}}$ and $M_{\text{CMV-}}$ are the number of patients with positive and negative CMV serostatus, respectively (different values for $M_{\text{CMV+}}$ and $M_{\text{CMV-}}$ are calculated for the training, validation, and test cohorts).

### *Mixing Conforming and Non-conforming Features*

Both the antigen and repertoire classification datasets contain conforming and non-conforming features that can be used in conjunction to classify each sample (Fig. 1c,d). For both datasets, separate weights are assigned for conforming and non-conforming features. The dot product of the conforming features and weights is computed separately and added to the dot product of the non-conforming features and weights computed using equation (2). The sum of the dot products represents the dot product of weights and combined features.

### *Scaling*

The dot product between the features and weights are computed piecewise for each group of features. For example, the dot product of the Vβ -gene features and associated weights is computed separately from the dot product of the CDR3-β features and associated weights. The weights of each feature group are scaled such that the dot product will have unit variance and zero mean, balancing the expected contribution from each feature group thereby implementing a naïve strategy of assigning equal importance to each feature group. The dot products from each feature group are added together to produce a dot product of all the features with all the weights. The weights are then scaled again to ensure the final dot product has unit variance and zero mean.

Scaling the weights is achieved using a procedure like batch normalization [4].

$$\mu = \sum_j f_j \cdot l_j \quad \sigma^2 = \sum_j f_j \cdot \left(l_j - \mu_j\right)^2 \quad l'_j = \frac{l_j - \mu}{\sqrt{\sigma^2}} \tag{1}$$

For any feature group, $l_j$ represents the value of the dot product (or output) from the $j^{\text{th}}$ sample. The sample frequencies $f_j$ weight the calculations (as defined in the previous section). The mean $\mu$ and variance $\sigma^2$ are calculated over the training dataset immediately after initializing the weights but before gradient optimization. Once $\mu$ and $\sigma^2$ are calculated they can be refactored into the weights and bias values (or left as an additional calculation after each dot product). This scaling operation is conceptually like batch normalization but applied over the entire dataset (not just a batch), weighted by sample frequency, and where $\mu$ and $\sigma^2$ are not updated during optimization.

### Similarities Between the Alignment Score and Max-Pooling

Both DKM and convolutional neural networks (CNNs) generate predictions from maximal responses. The alignment score in DKM represents the maximum possible sum of similarity scores between $\vec{x}_i$ and $\vec{\theta}_j$, whereas the pooling layer of CNNs represents the maximum possible responses from the convolutional layer. The reason why the alignment score works for classifying non-conforming features may be like the reason why pooling works in CNNs. The similarity of DKM with CNNs suggests that we can swap an alignment algorithm for each max-pooling operation in a multi-layer CNN to create a deep DKM neural network for classifying non-conforming features.

*Loss Function:* Numbers for the weights and bias are picked to minimize a loss function. We use the KL-divergence as the loss function, which measures the number of extra computer bits required to encode the labels when the predictions are used instead of the labels.

$$D_{KL} = -\sum_j f_j \cdot \sum_{i=1}^{M} y_j^{(i)} \cdot \ln{\left. p_j^{(i)} \middle/ y_j^{(i)} \right.}$$

$D_{KL}$ represents the KL divergence. The smaller the number the better the fit, with $0$ representing the best fit possible. The outer summation is over all samples in the datasets, indexed by $j$. Each sample is weighted by its frequency $f_j$, representing the relative number of times that sample appears in the dataset. The inner summation is over the label categories. $M$ represents the number of label categories and $i$ indicates a specific label category. For the antigen classification problem, $M = 6$ and for the repertoire classification problem $M = 2$. The labels (ground truth) for each category are represented by $y_j^{(i)}$ and the probabilities assigned to each category by the statistical classifier are represented by $p_j^{(i)}$.

### *Optimization*

We use adam, a gradient optimization-based technique, to minimize the loss function, referred to as the *fit*, over a cohort of samples used for training the statistical classifier [5]. To start, the weights are randomly *initialized*, which is where random numbers are assigned as the initial values for the weights. We use distributions described by Glorot and Bengio when drawing these random numbers [6]. Next, the optimization routine is repeatedly run for many steps. At every step, (i) equation (1) is used to compute the similarity scores, (ii) symbols are matched to maximize the alignment score as described, (iii) the alignment score is treated as a sum of features multiplied by weights, like in equation (2), (iv) a prediction is made, (v) the loss function is calculated by comparing predictions to labels, (vi) the negative of the gradient of the loss function with respect to the weights is computed, and (vii) the direction and magnitude of the gradients are used to make small changes to the weights to reduce the loss function. At each step, we must rerun i through vii.

Gradient optimization techniques do not guarantee finding the globally optimal solution, which is why we refit each statistical classifier multiple times, attempting to verify the best fit. With each refit, the weights are re-initialized before repeatedly running the optimization routine, allowing us to check that the fit does not depend on the initial conditions. When the fits vary considerably, we take the best fit *as measured over samples in the training cohort*, representing our attempt to find the global optimal fit (Supplementary Fig. 8j). Only the weights from the best fit are then used when evaluating samples from the validation and test cohorts.

### *Avoiding Overfitting*

With DKM, weights are assigned to features based on a similarity score rather than having a dedicated weight for each feature, decoupling the number of weights from the number of features. In this study, we pick the number of weights to be an order of magnitude less than the number of samples in the training cohort, reducing the risk of overfitting, a phenomenon where a statistical classifier fails to generalize to holdout samples.

## Supplementary Text, Results

### *Other Variations of DKM on the Repertoire Classification Problem*

To create a statistical classifier to handle the set of sequences, we represent the symbols in Θ as a set of sequences (see *Matching Features from a Set of Sequences to Weights* for the definition of sequences in Θ). For Θ, we need to pick the number of sequences in the set using our understanding of the problem to make an appropriate selection. Therefore, we turn to our understanding of immunology. In typical immune responses, only a handful of T-cells express a receptor that can bind the underlying pMHC. The remaining T-cells, representing most of the repertoire, are bystanders. To constrain the statistical classifier to identify the single most relevant TCR, ignoring receptors from bystander T-cells, we pick to have only one sequence in the set for Θ. We solve the assignment problem by brute force to identify the single most relevant CDR3 sequence in the TCR repertoire to match with the single sequence in the set for Θ.

In Supplementary Fig. 9, we show the result with this constraint relaxed, picking to have two sequences in the set for Θ. We again solve the assignment problem by brute force to identify two relevant CDR3 sequences in the TCR repertoire to match the two sequences in the set of Θ. With this added capacity, we observe the performance of the statistical classifier marginally improving.

### Analyzing Patient Age as a Feature on the Repertoire Classification Problem

Because elderly populations are more vulnerable to CMV infection and it is well known that there is a correlation between age and CMV positivity, we wondered if the statistical classifier is using each patient's age to predict their CMV serostatus. We included patient age to capture this effect in our model. Because of the simplicity of our model, we can look at the weight on the patient age to determine the importance of this feature in the model's predictions. Surprisingly, the weight for the patient age has a value of 0.00938 or almost zero. By comparison, the average magnitude for the weights on the Atchley factors used to represent each amino acid residue is 0.212. Therefore, patient age makes no practical contribution in the predictions of our statistical classifier. We do not mean to imply that age does not predict CMV serostatus, only that our statistical classifier was unable to uncover age as a predictor of CMV serostatus from the dataset used in this study.

### Capturing Additional Samples with Confidence Cutoffs

To capture additional samples, we can remove samples captured by the confidence cutoff, which are classified with an accuracy ≥95%, and refit the statistical classifier to the uncaptured samples, using the refitted model to capture additional samples with a classification accuracy of ≥95%. We can continue repeating this process to capture additional samples with an accuracy of ≥95% until no more samples can be captured with this degree of accuracy. The approach resembles boosting, where multiple weak statistical classifiers are combined into a single, stronger statistical classifier. We anticipate our future studies will combine DKM augmented statistical classifiers with boosting methods, like described, achieving higher classification accuracies over a larger subset of the samples.

## Supplementary Text, References

[1] J. Ostmeyer, S. Christley, W. H. Rounds, I. Toby, B. M. Greenberg, N. L. Monson and L. G. Cowell, "Statistical classifiers for diagnosing disease from immune repertoires: a case study using multiple sclerosis.," *BMC Bioinformatics,* vol. 18, no. 1, p. 401, 2017.

[2] J. Ostmeyer, S. Christley, I. T. Toby and L. G. Cowell, "Biophysicochemical motifs in T-cell receptor sequences distinguish repertoires from tumor-infiltrating lymphocyte and adjacent healthy tissue," *Cancer Research,* vol. 79, no. 7, pp. 1671-1680, 2019.

[3] J. L. Ostmeyer, L. G. Cowell and S. Christley, "Developing and validating an approach for diagnosing and prognosticating cancer from biochemical motifs in T-cell receptors.," *Journal of Clinical Oncology,* vol. 38, 2020.

[4] S. Ioffe and C. Szegedy, "Batch Normalization: Accelerating Deep Network Training by Reducing Internal Covariate Shift," in *Proceedings of The 32nd International Conference on Machine Learning*, 2015.

[5] D. P. Kingma and J. L. Ba, "Adam: A Method for Stochastic Optimization," in *ICLR 2015 : International Conference on Learning Representations 2015*, 2015.

[6] X. Glorot and Y. Bengio, "Understanding the difficulty of training deep feedforward neural networks," in *Proceedings of the Thirteenth International Conference on Artificial Intelligence and Statistics*, 2010.